\documentclass[twocolumn,showpacs,preprintnumbers,floatfix,amsmath,amssymb,
superscriptaddress]{revtex4}

\usepackage{graphicx}
\usepackage{dcolumn}
\usepackage{bm}

\newcommand{\be}{\begin{equation}}
\newcommand{\ee}{\end{equation}}
\newcommand{\bn}{\begin{eqnarray}}
\newcommand{\en}{\end{eqnarray}}
\newcommand{\ba}{\begin{array}}
\newcommand{\ea}{\end{array}}
\newcommand{\bc}{\begin{center}}
\newcommand{\ec}{\end{center}}
\newcommand{\bml}{\begin{mathletters}}
\newcommand{\eml}{\end{mathletters}}

\begin{document}


\title{Contradicting effective mass scalings within the Skyrme
energy density functional method}

\author{W. Satu{\l}a}
\email{             satula@fuw.edu.pl}
\affiliation{Institute of Theoretical Physics, University of Warsaw, ul. Ho\.za
69, 00-681 Warsaw, Poland}%
\affiliation{          KTH (Royal Institute of Technology),\\
           AlbaNova University Center, 106 91 Stockholm, Sweden}

\author{R. A. Wyss}
\email{           wyss@kth.se}
\affiliation{          KTH (Royal Institute of Technology),\\
           AlbaNova University Center, 106 91 Stockholm, Sweden}

\author{M. Zalewski}
\email{             zalewiak@fuw.edu.pl}
\affiliation{Institute of Theoretical Physics, University of Warsaw, ul. Ho\.za
69, 00-681 Warsaw, Poland}

\date{\today}

\begin{abstract}
The problem of the effective mass scaling in the single particle
spectra calculated within the Skyrme
energy density functional (EDF) method is studied. It is demonstrated
that for specific pairs of orbitals the commonly anticipated isoscalar
effective mass ($m^*$) scaling
of the single-particle level splittings is almost canceled
by an implicit $m^*$-scaling due to other parameters in
the Skyrme EDF. This holds in particular for an indirect $m^*$-scaling
of the two-body spin-orbit strength making the theory
essentially unpredictable with respect to single particle
energies. It is argued that this unphysical property
of the Skyrme EDF is a mere consequence
of the strategies and datasets used to fit these functionals.
The inclusion
of certain single-particle spin-orbit splittings to fit
the two-body spin-orbit and the tensor interaction strengths
reinstates the conventional $m^*$-scaling and improves the
performance of the Skyrme EDF.
\end{abstract}

 \maketitle

Fundamental excitations in atomic nuclei are often characterized in terms of single
particle and collective excitations. A successful nuclear theory is required
to account for both kinds at a qualitative {\sl and}
quantitative level.  The effective nuclear energy density functionals (EDF)
irrespectively of their
variants including local (Skyrme), non-local (Gogny) or relativistic
mean-field (RMF) have developed toward high accuracy in recent years, both
with respect to bulk properties and
predictions of single particle properties\cite{[Sat05]}.
Still, on a quantitative level,
properties of single particle states are often better described by means
of simple potentials like the Nilsson or Woods-Saxon. This is highly
unsatisfactory, since one expects
a full fledged EDF to describe nuclear properties at a level of accuracy superior to
that of simple potentials. The uncertainty of present EDF with respect to nuclear
properties has its roots in the fitting procedure of its parameters as well as
the non local terms in the interaction.

The EDF's 
are conventionally adjusted to reproduce ground-state bulk
nuclear properties, see e.g. Ref.\cite{[Cha97w]}.
The datasets used to fit their parameters are
dominated by nuclear data extrapolated to the thermodynamic limit and by
nuclear binding energies in selected doubly magic nuclei.
These fitting procedures are known to impair basic buildings blocks
of these theoretical models, in particular single-particle (s.p.) energies.
The physical relevance of s.p. energies provided by
self-consistent mean-field (MF) approaches based on the EDFs or effective
interactions is continuously contested. The debate has its roots in the
non-locality of these approaches resulting in a low isoscalar effective mass
$\frac{m^*}{m}\approx 0.8$ \cite{[Jeu76w]} which in turn scales the s.p. level
density $g$ in the vicinity of the Fermi energy $\varepsilon_F$ according to the
simple rule: $g(\varepsilon_F) \rightarrow \frac{m}{m^*}g(\varepsilon_F)$.
This simple rule applies strictly to homogeneous nuclear matter only. In finite nuclei
the effective mass depends on $\boldsymbol r$ and the
 $g(\varepsilon_F) \rightarrow \frac{m}{m^*}g(\varepsilon_F)$ scaling
should be considered as an idealization.
The effective mass scaling leads to a dichotomous and in fact highly
uncomfortable situation. Indeed, in spite of the fact that it makes all
applications of the self-consistent MF methods to low-lying nuclear
excitations rather dubious such calculations are carried out routinely and
published often without even a word of comment or justification.

\smallskip

The aim of this Letter is to demonstrate that the problem of the
effective mass scaling within the {\it effective theory\/}
is far more intricate than anticipated. It
turns out that the effective mass dependence of the calculated
single particle spectrum can depend on the structure of the EDF,
on the strategies employed
when adjusting the interactions and upon the choice of dataset used in these fits.
We will demonstrate first that conventional functionals which
are, as discussed above, fitted almost ultimately to bulk nuclear properties
have an built in {\it implicit\/} effective mass scaling of certain coupling constants
including in particular the two-body spin-orbit strength.
For many modern parameterizations of the Skyrme force this
mechanism is strong enough to cancel the
effect of the s.p. level density scaling caused by
the low effective mass with respect to
the calculated specific particle-hole excitation.
For example the $d_{3/2}-f_{7/2}$ splitting
in $A$$\sim$44 mass region~\cite{[Zdu05]} calculated using forces having
effective masses ranging from $0.7\leq \frac{m^*}{m}\leq 1$ yield
similar result in spite of the
expected effective mass dependence.
In the next step we will show
that this counterintuitive result is a mere consequence of the fitting
strategies and that by
shifting the attention from bulk to
single-particle properties in the process of fitting of the nuclear EDF
parameters one can both remove the artificial $m^*$-scaling from
the two-body spin-orbit strength and reinstate the conventional
and anticipated $m^*$-scaling in the calculated $ph$ excitation energies.

\smallskip

All the calculations performed in this Letter are based on
Skyrme-force inspired local EDF (SEDF) of the form:
\begin{equation}\label{sky}
\mathcal{H}({\boldsymbol r}) = \sum_{t=0,1} \left(
\mathcal{H}_t^{\text{even}}({\boldsymbol r}) +
\mathcal{H}_t^{\text{odd}}({\boldsymbol r}) \right) \, ,
\end{equation}
where
\begin{eqnarray} \label{hte}
\mathcal{H}_t^{\text{even}} & = & C^{\rho}_t
\rho^2_t + C^{\Delta \rho}_t \rho_t\Delta\rho_t + \\ \nonumber &\quad &
C^{\tau}_t \rho_t\tau_t + C^J_t {\mathbb J}^2_t + C^{\nabla J}_t \rho_t
{\mathbf \nabla}\cdot{\mathbf J}_t,
\end{eqnarray}
\begin{eqnarray}
\label{hto} \mathcal{H}_t^{\text{odd}} & = &C^{s}_t {\mathbf s}^2_t +
C^{\Delta s}_t {\mathbf s}_t\cdot\Delta {\mathbf s}_t +  \\ \nonumber &\quad
&C^{T}_t{\mathbf s}_t \cdot {\mathbf T}_t + C^j_t {\mathbf j^2_t} + C^{\nabla
j}_t {\mathbf s}_t \cdot ({\mathbf \nabla}\times {\mathbf j}_t).
\end{eqnarray}
It depends on isoscalar $t=0$ and
isovector $t=1$ time-even $\rho_t$, $\tau_t$, and ${\mathbb J}_t$,
and time-odd ${\mathbf s}_t$, ${\mathbf T}_t$, and ${\mathbf j}_t$, local
densities for which we follow the convention introduced
in Ref.~\cite{[Eng75s]}, see also Refs.~\cite{[Ben03],[Per04s]} and references
cited therein. All numerical results presented below were obtained using the
HFODD code of Ref.~\cite{[Dob04w]} and the following set of the Skyrme
forces: SLy4$_L$, SLy5$_L$~\cite{[Cha97w]}, SIII$_L$~\cite{[Bei75s]},
SkM$^*_L$~\cite{[Bar82s]}, SkXc$_L$~\cite{[Bro98w]}, MSk1$_L$~\cite{[Ton00s]},
SkP$_L$~\cite{[Dob84w]}, and SkO$_L$~\cite{[Rei99s]}. The subscript
$L$ indicates here that the original time-odd functional coupling constants
$C^s, C^{\Delta s}$, and $C^T$ were replaced in the calculations by
the coupling constants reproducing the empirical Landau parameters in
accordance to the prescription given in Refs.~\cite{[Ben02s],[Zdu05]}.

\smallskip

In spite of the fact that we will concentrate here on the {\it implicit\/}
$m^*$-scaling of certain functional coupling constants it should be mentioned
that some of the SEDF parameters do scale or depend upon  $m^*$ explicitly. The
{\it explicit\/} effective mass scaling is well established for the ({\it i\/})
$C^s$ and $C^T$ functional parameters through the fit to the empirical Landau
parameters~\cite{[Ben02s],[Zdu05]} and ({\it ii\/}) for the isovectorial
coupling constants $C^\rho_1$ and $C^\tau_1$ through the fit to the empirical
symmetry energy strength~\cite{[Sat06w]}.

\vspace{0.3cm}

An almost ideal playground to investigate the
impact of fitting procedures on the performance of the effective forces
with respect to the $ph$ related observables is offered by fully
stretched $[f_{7/2}^n]_{I_{max}}$ and $[d_{3/2}^{-1} f_{7/2}^{n+1}]_{I_{max}}$
states in $N$$\ne$$Z$ nuclei in the $A$$\sim$44 mass region
where $n$ denotes number of valence particles in the $f_{7/2}$ sub-shell
\cite{[Zdu05]}. In a series of
publications~\cite{[Zdu05],[Sto06s],[Zal07s],[Zdu07s]}
we have
demonstrated that these states represent one of the best examples of
almost unperturbed single-particle motion and that
the energy difference:
\begin{equation}\label{de}
\Delta E = E([f_{7/2}^n]_{I_{max}}) - E([d_{3/2}^{-1}
f_{7/2}^{n+1}]_{I_{max}})
\end{equation}
constitutes a very reliable probe of
various properties of the EDF. In particular, it can be used to
readjust the time-odd components $C^s$, $C^{\Delta s}$ and $C^T$
to comply with the empirical Landau parameters leading to
an unification of the theoretical predictions for $\Delta E_{th}$
for various popular Skyrme parameterizations~\cite{[Zdu05]}.

%
\begin{figure}[t]
\centerline{\includegraphics[
width=0.45\textwidth,clip]{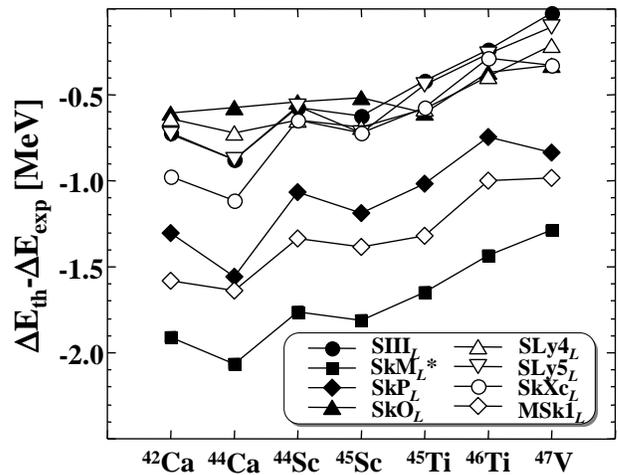}}
\caption{The difference between theoretical and empirical values of
the energy differences $\Delta E$ defined in
Eq.~(\protect{\ref{de}}). Calculations have been done using
different popular parameterizations (see legend) of the Skyrme
functionals with spin-dependent
coupling constant readjusted to reproduce the empirical
Landau parameters in accordance to
Refs.~\protect{\cite{[Ben02s],[Zdu05]}}.
}
\label{ttf1}
\end{figure}

The calculated values of  $\Delta E_{th}$, see Fig.~\ref{ttf1},
show a striking and completely unexpected feature. The
mean deviation of the theoretical predictions with respect to the
empirical values is very similar for such forces like SLy4, SLy5
($\frac{m^*}{m}\sim 0.7$), SIII ($\frac{m^*}{m}\sim 0.8$),
SkO  ($\frac{m^*}{m}\sim 0.9$) and SkXc  ($\frac{m^*}{m}\sim 1.0$).
In spite of the fact the the effective masses differ by as much as 30\%,
 we obtain for these parameterizations
$\delta {\bar E} = \Delta {\bar E}_{th}-\Delta {\bar
E}_{exp} \sim - 550$\,keV with a rms deviation of $\sigma \approx 70$\,keV. This
result is indeed extremely puzzling since the anticipated
influence of a naive $m^*$-scaling  with respect to a $ph$ excitation
energy of order of $\Delta {\bar E}_{exp} \approx 5.5$\,MeV
is, for the analyzed set of forces, estimated to be more than one order of
magnitude larger, exceeding $\sigma \approx 1$\,MeV.

\vspace{0.3cm}

\begin{figure}[t]
\centerline{\includegraphics[
width=0.45\textwidth,clip]{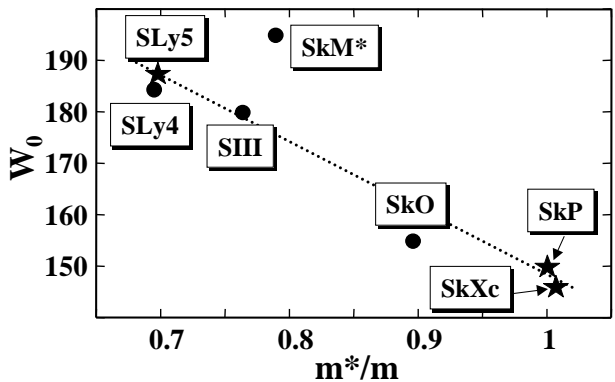}}
\caption{The isoscalar strength of the two-body spin-orbit interaction
plotted as a function of the isoscalar effective mass parameter for
a few popular Skyrme force parameterizations studied here including
SLy4, SLy5 SkO, SIII, SkXc, SkP, and SkM$^*$. Stars mark
the Skyrme forces having non-zero tensor terms.
}
\label{ttf2}
\end{figure}

Unexpectedly, the observable $\Delta E$ appears to be very robust. Indeed,
the structural purity of the terminating states in the $A$$\sim$44
mass region reveal a firmly established hierarchy of
different physical contributions to that quantity
and in turn allows to identify the physical source of the {\it
cancellation\/} of the $m^*$-scaling in the functional. In
Ref.~\cite{[Zdu05]} we established the hierarchy of three different components,
the energy scale $\hbar\omega$, the strength of the spin-orbit (SO) and the
$\ell^2$ (surface) term using the Nilsson Hamiltonian.
This schematic model
clearly indicates  the dominant influence of the mean
SO potential
on $\Delta E$. This observation is neatly correlated with
self-consistent SHF models, see the Appendix below where we give a
numerical proof for the
case of the Skyrme functionals. Indeed, the isoscalar strength of the Skyrme
one-body SO potential emerging from the two-body SO interaction
$W_0=-2C_0^{\nabla J}$ appears to be almost perfectly linearly
correlated with the effective mass $m^*$ parameter
as shown in Fig.~\ref{ttf2}. Large values of
$W_0$ which characterize low$-m^*$ forces tend to reduce
the $d_{3/2}$-$f_{7/2}$ splitting thus compensating the
$m^*$-scaling effect on the $N$=$Z$=20 magic gap. As shown in
Ref.~\cite{[Zal07w]} a similar cancellation takes place
for the $p_{1/2}$-$d_{5/2}$ and  $f_{5/2}$-$g_{9/2}$ splittings in
the $A$=16 and $A$=80 mass regions, respectively. For
these splittings the schematic Nilsson model indicates an indisputable
dominance of the SO term over the surface
(or, more precisely, the $\sim \ell^{\,2}$) contribution.
On the other hand, for the $p_{1/2}$-$g_{9/2}$ splitting in the $A$=80 mass region,
the SO and the $\ell^{\, 2}$ contributions
are predicted to be similar. For this particular case also the
self-consistent SHF calculations provide
results which are qualitatively different than in the previous
cases indicating a more complex dependence.

%


%
\begin{figure}[t]
\centerline{\includegraphics[
width=0.45\textwidth,clip]{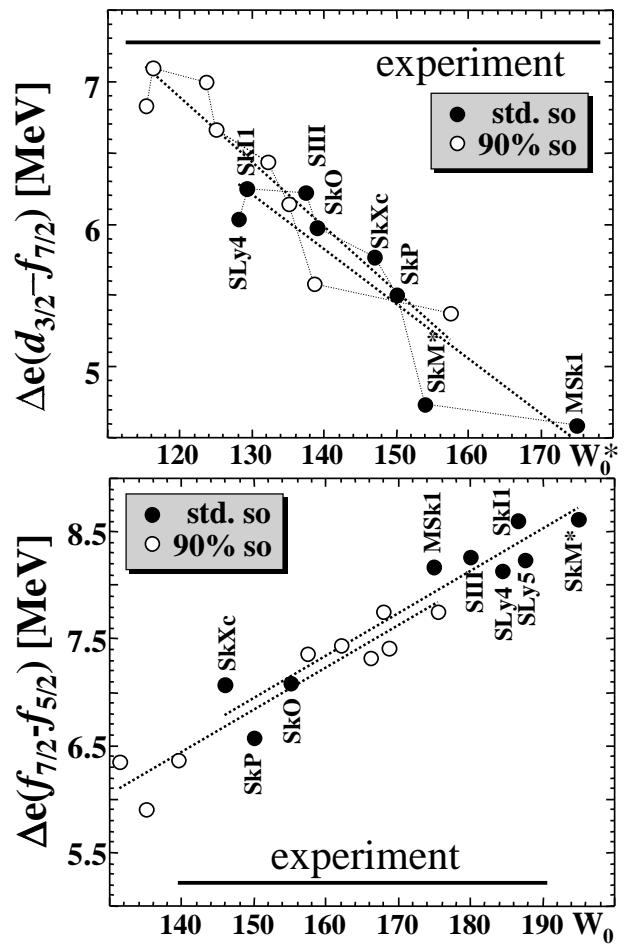}}
\caption{Upper part shows the calculated energy splitting between
$\nu f_{7/2}-\nu d_{3/2}$ single-particle states
versus $W^*_0$ calculated using different Skyrme forces. The
lower part shows  the $\nu f_{7/2}-\nu f_{5/2}$
spin-orbit splitting versus $W_0$ calculated using
the same set of the Skyrme forces.
Black (open) dots mark results obtained using standard
(10\% reduced) strength of the two-body spin-orbit
interaction, respectively. The figure clearly demonstrates
the presence of two conflicting scalings within the Skyrme model.
See text for further details.
}
\label{ttf3}
\end{figure}

The {\it implicit\/} mass scaling of the two-body Skyrme
spin-orbit strength has unexpected and serious consequences:
Within a single theoretical
framework two conflicting scalings are
present.
Indeed, particle-hole excitations
associated with the spin-orbit partners such as $f_{7/2}-f_{5/2}$ scale
directly with $W_0$. At the same time the $f_{7/2}-d_{3/2}$ $ph$ excitations
scale according to the $W_0^* \equiv \frac{m^*}{m}W_0$ as shown in
Fig.~\ref{ttf3}. At first glance this puzzling situation seem to have no
satisfactory and unique solution within the conventional
Skyrme EDF. Although a reduction in $W_0$ (see open symbols in Fig.~\ref{ttf3})
 clearly improves the agreement to the data it
can hardly be accepted as a reasonable solution. Indeed,
the figure indicates that the empirical $\Delta e(f_{7/2}-f_{5/2})$ splitting
(experimental data are marked by thick horizontal lines in Fig.~\ref{ttf3})
is reached first by reducing $W_0$ strength in parameterizations
characterized by large-${m^*}$ values.
In contrast, agreement to the empirical $\Delta e(f_{7/2}-d_{3/2})$ splitting is
obtained after reducing the $W_0$ strength in
low-${m^*}$ parameterizations. Therefore, a
conventional theory can
not reproduce simultaneously both empirical values.
To state it differently: the standard set of Skyrme forces appear to be incomplete
when confronted to experimental data.

 \vspace{0.3cm}

Fig.~\ref{ttf3} suggests that the simultaneous agreement for
both  $\Delta e(f_{7/2}-d_{3/2})$  and   $\Delta e(f_{7/2}-f_{5/2})$
splittings calls for parameterizations having large
${m^*}$ and drastically reduced $W_0$. Drastic
reduction of $W_0$ will spoil, however, the relatively good agreement for our
high-spin observable $\Delta E$, see Fig.~\ref{ttf1}. Hence a compensation mechanism
is required. This mechanism exist, it is not
new~\cite{[Flo75w],[Sta77w]} and it is provided by a  tensor
component.
However, until very recently, there has been little
need to invoke the {\sl strong} tensor interaction in EDF's,
see~\cite{[Ots01s],[Ots05s],[Dob06s],[Bro06s],[Les07s],[Col07s],[Gra07s]}.
In our recent contribution~\cite{[Zal08s]}
we have shown that the tensor interaction can be rather unambiguously fitted using the
$f_{7/2}-f_{5/2}$ spin-orbit splittings in three key nuclei including:
the isoscalar spin-saturated $^{40}$Ca nucleus, the isoscalar spin-unsaturated
$^{56}$Ni nucleus, and the isovector spin-unsaturated $^{48}$Ca nucleus.
Unlike in the other studies cited above our work firmly revealed
the need for a simultaneous drastic reduction of the two-body SO strength.
Such a procedure definitely changes
the philosophy behind conventional fitting strategies by shifting the attention
from mass dominated gross features to procedures including carefully selected
single-particle states which are used to adjust specific coupling
constants in the functional.

\smallskip

We are now in a position to verify the consistency of
our new fitting strategy using  high spin terminating
states. We have therefore repeated the
calculations for the energy differences $\Delta E$ of Eq.~(\ref{de})
but using parameterizations SIII$_{LT}$, SkO$_{LT}$, SLy4$_{LT}$,
SkP$_{LT}$, and SkXc$_{LT}$.  Subscript $LT$ indicates that these parameterizations
have: ({\it i\/}) time-odd spin-fields readjusted to empirical Landau
parameters and ({\it ii\/}) modified tensor and two-body SO strengths. New
values of the tensor and the two-body SO functional coupling constants are
collected in Tab.~\ref{fity}. All other coupling constants of these
parameterizations are kept to their original values.

%
\begin{figure}[t]
\centerline{\includegraphics[
width=0.45\textwidth,clip]{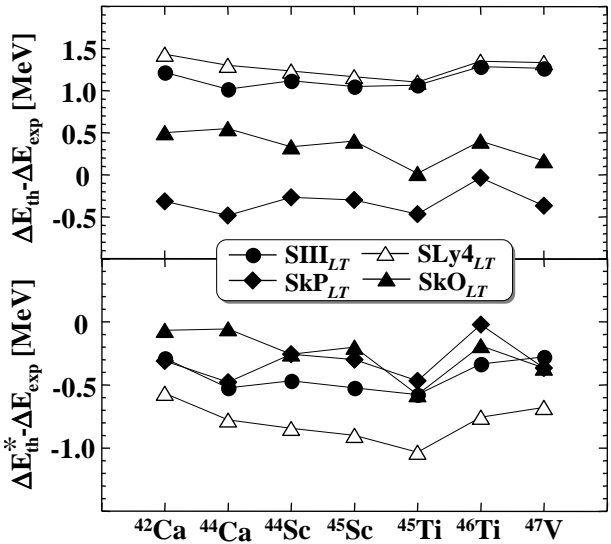}}
\caption{The energy difference of theoretical and empirical $\Delta E_{th} -
\Delta E_{exp}$ energy splittings given by Eq.~(\protect{\ref{de}}). Upper
part shows the results obtained using SLy4$_{LT}$, SkP$_{LT}$, SIII$_{LT}$,
and SkO$_{LT}$ parameterizations with modified spin
fields and the two-body SO and tensor strengths readjusted to the empirical
$\nu f_{7/2}-\nu f_{5/2}$ splittings, see table~\protect{\ref{fity}}.
Lower port shows the differences $\Delta E_{th}^* -
\Delta E_{exp}$ between the isoscalar effective mass
scaled theoretical splitting $\Delta E_{th}^* =
\frac{m^*}{m}\Delta E_{th}$ and the experimental value.
See text for further details.}
\label{ttf4} \end{figure}

\begin{table}[t]
\begin{center}
\begin{tabular}{ccccc}
\hline
\hline
Skyrme &  $C_{0}^{\nabla J}$ &    $C_{1}^{\nabla J}/C_{0}^{\nabla J}$   &
$C_{0}^J$  & $C_{1}^J$ \\
 force &   [MeV\,fm$^5$]          &    & [MeV\,fm$^5$]    &  [MeV\,fm$^5$]  \\
\hline
\hline
SkP$_T$    &     -60.0     &  1/3    &    -38.6      &    -61.7            \\
SLy4$_T$   &     -60.0     &  1/3    &    -45.0      &    -60.0            \\
SIII$_T$   &     -57.6     &  1/3    &    -50.6      &    -64.5            \\
\hline
SkO$_T$    &     -61.8     & -1.3    &    -33.1      &    -91.6            \\
SkXc$_T$   &     -54.0     &   0     &    -43.0      &    -65.2            \\
\hline
\hline
\end{tabular}
\caption[A]{Spin-orbit $C^{\nabla J}$ and tensor isoscalar $C_{0}^J$  and
isovector $C_{1}^J$ functional coupling constants
adopted for different parameterizations. These
modified parameterizations are subsequently used in the
calculations presented in Fig.~\ref{ttf4}. }
\label{fity} \end{center}
\end{table}

%
\begin{figure}[t]
\centerline{\includegraphics[
width=0.45\textwidth,clip]{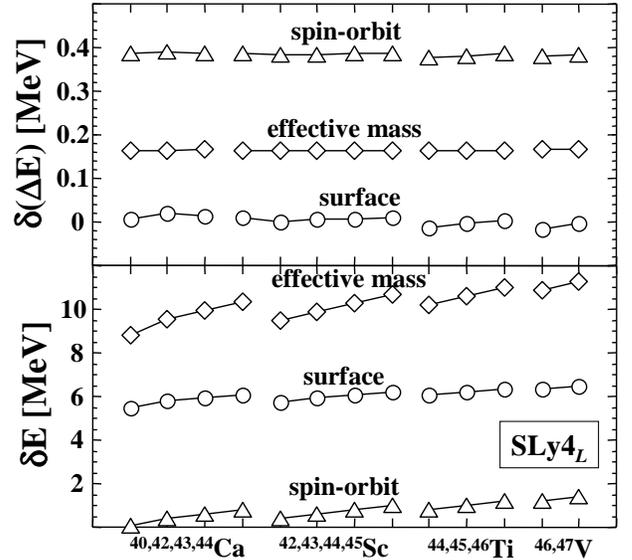}}
\caption{The change in $\Delta E$ (upper part) and in the binding
energy $E([f_{7/2}^n]_{I_{max}})$ (lower part) caused by 5\% variations in
$C^\Delta\rho$ (surface), $C^{\tau}$ (effective mass),
and $C^{\nabla J}$ (spin-orbit) coupling constants in the
Skyrme functional.}
\label{ttf5}
\end{figure}

The results of our calculations
are depicted in the upper part of Fig.~\ref{ttf4}. It is interesting
to observe that all large-$m^*$ parameterizations
including SkO$_{LT}$ ($\frac{m^*}{m}\sim 0.9$) and
SkP$_{LT}$ ($\frac{m^*}{m}\sim 1.0$) are giving satisfactory
agreement to the data. In the particular case of the
SkXc$_{LT}$ ($\frac{m^*}{m}\sim 1.0$) force (not shown in the figure)
the mean deviation from the data equals to $33$\,keV only!
At the same time low-$m^*$ parameterizations
including SLy4$_{LT}$ ($\frac{m^*}{m}\sim 0.7$) and
SIII$_{LT}$ ($\frac{m^*}{m}\sim 0.8$) strongly overestimate the
data. This result clearly follows the conventional wisdom
related to the $m^*$-scaling of the s.p. level density.
To visualize this even better we have rescaled the theoretical energy
difference by the effective mass $\Delta E_{th}^* \equiv
\frac{m^*}{m}\Delta E_{th}$ and have plotted the difference
$\Delta E_{th}^* - \Delta E_{exp}$ in the lower panel of Fig.~\ref{ttf4}.
Three out of four curves depicted in this figure follow closely
each other and the experiment. The curve representing the SLy4$_{LT}$ is
slightly below the trend, indicating too small a reduction
of the two-body SO strength  (by 35\%) done in Ref.~\cite{[Zal08s]}.
The quantity $\Delta E$ can therefore be used for further
fine tuning of the SO strength and indeed, a reduction of the SO strength by
40\% shifts the SLy4 curve by 250keV up as expected. One has to remember
however, that the entire effective mass scaling concept in finite nuclei is
only an idealization. Last but not least let us observe that all curves shown
in Fig.~\ref{ttf4} do not reveal any clear isotopic and/or isotonic
dependence, in contrast to the previous calculations presented in
Fig.~\ref{ttf1}. This indicates a clear improvement of the isovector channel
due to the presence of a strong isovector tensor component in the functional.

\smallskip

In summary, we have investigated the impact of the fitting procedures
and the datasets used to fit the Skyrme energy density functionals
with respect to their spectroscopic properties. We have demonstrated that the
use of parameterizations fitted to reproduce bulk
nuclear data in the thermodynamic limit and to binding energies
of selected double-magic finite nuclei to analyze
spectroscopic data may lead to rather poor
results. This is due to an implicit internal $m^*$-scaling
of, in particular, the two-body SO strength which is a mere
consequence of the fitting procedure.
We have further demonstrated that conventional Skyrme forces
having the two-body SO strength and tensor coupling constants
fitted directly to the empirical s.p. $\nu f_{7/2}- \nu f_{5/2}$
SO splittings behave according to the expected $m^*$-scaling
law for particle-hole excitations.
The present study seems to confirm the
common expectation that
effective interactions with large effective mass
have a superior performance for calculations of spectroscopic data.
It sends however a clear message that this conclusion is strongly
dependent on the fitting process for effective forces. Involving
a larger set of single particle data, in particular selected
high spin terminating
states reveal the
need for a considerable reduction of the conventional
two-body spin-orbit term and at the same time the requirement for
a strong tensor term introduced
in Ref.~\cite{[Zal08s]}.


\vspace{0.3cm}

This work was supported in part by the Polish Ministry of
Science and the Swedish Research Council.

\vspace{0.3cm}
\section{
APPENDIX}

In order to demonstrate the hierarchy of
various contributions to the observable (\ref{de})
in $A$$\sim$44 mass region we have performed
additional calculations using the Skyrme EDF~(\ref{hte})-(\ref{hto}).
The results are depicted in Fig.~\ref{ttf5}. The
upper part of the figure shows relative changes $\delta(\Delta E)$ in
the energy differences $\Delta E$ caused by $\pm$5\% changes in
$C^{\Delta\rho}$ (surface), $C^{\tau}(=-C^j)$ (effective mass), and
$C^{\nabla J}(=C^{\nabla j})$ (spin-orbit) coupling constants with respect
to their original values. The lower part illustrates
relative changes in the total binding energy $\delta E([f_{7/2}^n]_{I_{max}})$
of the aligned state $[f_{7/2}^n]_{I_{max}}$ caused by
these variations in the coupling constants. This figure
clearly shows that the hierarchy established in Ref.~\cite{[Zdu05]}
on the basis of the schematic Nilsson model is
correct. The leading contribution to $\delta(\Delta E)$ is indeed the
spin-orbit term. The contribution coming from the effective mass term is circa 2.5
times smaller and is clearly non-perturbative as it impacts binding energies
by $\sim$10\,MeV. In the same spirit we have also investigated the influence
of $\pm$1\% variations in the bulk energy parameter $C^{\rho}$
on both $\Delta E$ and $E([f_{7/2}^n]_{I_{max}})$.
These tiny variations in $C^{\rho}$ impact binding energies by
$\sim$15\,MeV and, at the same time, affect
 $\Delta E$ only by $\sim$$\pm$30\,keV.


\end{document}